\documentclass[aps,preprint,showpacs,nofootinbib,12pt]{revtex4}
\usepackage{graphicx}
\usepackage{epsfig}
\usepackage[dvips]{color}

\begin{document}

\title{Possible molecular states of $\bar D^{*}K^{*}$ ($ D^{*}K^{*}$) and the new exotic states
 $X_0(2900)$ and $X_1(2900)$ ($T^a_{cs0}(2900)^0$ and $T^a_{cs0}(2900)^{++}$) }

\author{ Hong-Wei Ke$^{1}$   \footnote{khw020056@tju.edu.cn}, Yi-Fan Shi$^{1}$, Xiao-Hai Liu$^{1}$ \footnote{xiaohai.liu@tju.edu.cn}  and
         Xue-Qian Li $^{2}$  \footnote{lixq@nankai.edu.cn}}

\affiliation{
  $^{1}$ School of Science, Tianjin University, Tianjin 300072, China \\
  $^{2}$ School of Physics, Nankai University, Tianjin 300071, China
 }

\begin{abstract}
\noindent

Two iso-singlet hadron states $X_0(2900)$ and $X_1(2900)$ with $J=0$ and 1 respectively, discovered by the LHCb collaboration in 2020,
are identified as molecular bound states of $\bar D^*K^*$. Recently two structures $T^a_{cs0}(2900)^0$ and $T^a_{cs0}(2900)^{++}$  have
been observed at the hadron spectra, and one would suspect if they also are molecular states of $D^*$ and $K^*$. As long as they are of the molecular structures of
$D^*K^*$, the hadron states must be in an iso-vector, namely $T^a_{cs0}(2900)^0$ and $T^a_{cs0}(2900)^{++}$ are $I_3=-1, 1$ components of the iso-vector.
If it is the case, the corresponding $T^a_{cs0}(2900)^+$ of $I=1,I_3=0$ and $T^{'a}_{cs0}(2900)^{+}$ of $I=0,I_3=0$ so far have evaded experimental observation, but
should be found by the future experiments. To testify this ansatz, in
this paper we study the possible molecular structures of $\bar
D^{*}K^{*}$ and $D^{*}K^{*}$ within the Bethe-Salpeter (B-S)
framework. With reasonable input
parameters it is found that $\bar D^{*}K^{*}$ iso-scalar systems with $J^P=0^+$ and $1^+$
have solutions. The result supports the ansatz of $X_0(2900)$ and
$X_1(2900)$ being molecular states of $\bar D^*K^{*}$. Whereas for the
system of $ D^{*}K^{*}$  with $I=1$ the corresponding B-S equation
has no solution. Thus we can draw a clear conclusion that
$T^a_{cs0}(2900)^0$ and $T^a_{cs0}(2900)^{++}$ should not be bound states of $ D^{*}$ and $K^{*}$. The two structures observed
by the LHCb collaboration may be caused by dynamics, such as the well-recognized triangle anomalies or other mechanisms.

\end{abstract}


\maketitle

\section{introduction}
In 2020 the LHCb Collaboration reported two exotic states
$X_0(2900)$ and $X_1(2900)$ in the $D^-K^+$ invariant mass
spectrum. Naturally the two states possess a $\bar c \bar s u
d$ flavor structure. Very recently the LHCb Collaboration claimed that
two new hadronic states $T^a_{cs0}(2900)^0$ and
$T^a_{cs0}(2900)^{++}$ in the mass distributions of
$D_s^+\pi^-$ and $D_s^+\pi^+$ had been observed. It strongly implies that the two states have  $c \bar s \bar
u d$ or $c \bar s u \bar d$  flavor structures.

The masses and widthes of the $X_0(2900)$, $X_1(2900)$ and $T^a_{cs0}(2900)$ are

\begin{eqnarray*}
&&m_{_{X_0(2900)}}=(2866\pm7)\, {\rm MeV},
\nonumber\\&&\Gamma_{_{X_0(2900)}}=(57.2\pm12.9)\, {\rm MeV},
\nonumber\\&&m_{_{X_1(2900)}}=(2904\pm5)\, {\rm MeV},
\nonumber\\&&\Gamma_{_{X_1(2900)}}=(110.3\pm11.5)\, {\rm MeV},
\nonumber\\&&m_{_{T^a_{cs0}(2900)}}=(2908\pm11\pm2)\, {\rm MeV},
\nonumber\\&&\Gamma_{_{T^a_{cs0}(2900)}}=(136\pm23\pm11)\, {\rm
MeV}.
\end{eqnarray*}

Those states are composed with four different flavor quarks (or antiquarks),
thus they are distinct from the previously confirmed exotic states such as
$X(3872)$, $X(3940)$, $Y(3940)$, $Z(4430)^{\pm}$, $Z_{cs}(4000)$,
$Z_{cs}(4220)$, $Z_b$, $Z_b'$,
$T_{cc}^+$ \cite{LHCb:2021uow,Choi:2007wga,Choi:2005,Abe:2007jn,Choi:2003ue,Collaboration:2011gj,
LHCb:2021auc,LHCb:2021vvq}.

Since $X_0(2900)$, $T^a_{cs0}(2900)^0$ and $T^a_{cs0}(2900)^{++}$
are made of $\bar c \bar s u d$,  $c \bar s \bar u d$ and $c \bar
s u \bar d$ respectively, they cannot be identified as the
traditional mesons  which contain a quark and an anti-quark pair.
Instead,  they are suggested to be multi-quark exotic states which were
predicted at the early stage when the $SU(3)$ quark model \cite{GellMann:1964nj} was raised. In fact, they might be
molecular states, compact
tetraquarks, mixing of both structures or non-hadronic resonances occurring due to dynamical
effects \cite{Chen:2016spr,Guo:2019twa}. To confirm their structures, one needs to invoke
a synthesis of theoretical analysis and experimental observation.

Since the masses of
$X_0(2900)$ and $X_1(2900)$  are near the threshold of $\bar D^*$ and $K^*$, they are identified as $\bar D^*K^*$ molecules. Similarly, as the masses
of $T^a_{cs0}(2900)^0$ and $T^a_{cs0}(2900)^{++}$ are close to
that of $D^*$ and $K^*$, one is tempted to expect these new states may be molecular
states of $D^*K^*$.
In some works the possible $\bar D^*K^*$  or $D^*K^*$ molecular states were proposed and studied \cite{Molina:2010tx,Molina:2020hde,Liu:2020nil,He:2020btl,Agaev:2020nrc}.

Our  starting point is that as long as $T^a_{cs0}(2900)^0$ and $T^a_{cs0}(2900)^{++}$
are molecular states composed of $c\bar sd\bar u$ and $c\bar su\bar d$ respectively, they would be
the ($I=1,I_3=-1)$ and $(I=1,I_3=1$) components of an iso-vector. If it is the case,  $T^a_{cs0}(2900)^+$
of ($I=1,I_3=0)$ corresponding to $c\bar s(u\bar u-d\bar d)$ and  $T^{'a}_{cs0}(2900)^+$ of $(I=0,I_3=0)$ for
$c\bar s(u\bar u+d\bar d)$ should exist and so far evade from experimental observation yet. We will probe this ansatz
in this work.

Concretely, we study the possible
bound state of $\bar D^*K^*$  and $D^*K^*$ systems within the
Bethe-Salpeter (B-S) framework where the relativistic corrections
are automatically included \cite{Guo:1998ef,Weng:2010rb,Li:2019ekr,Ke:2012gm,Ke:2020eba,Ke:2019bkf,Guo:2007mm,Feng:2011zzb,Ke:2018jql,Feng:2012zzf,Ding:2021igr,Ke:2021iyh,Ke:2021rxd,Ke:2022vsi,Oller:2000ma}.

In this work we employ the one-boson-exchange model to obtain
the interaction kernels where the effective vertices  ($D^*D^* M$
and $\bar D^*\bar D^* M$) are taken from the heavy meson chiral
perturbation
theory \cite{Colangelo:2005gb,Colangelo:2012xi,Ding:2008gr,Casalbuoni:1996pg,Casalbuoni:1992gi,Casalbuoni:1992dx}
and the effective vertices ( $K^*K^* M$) are adopted from Ref. \cite{He:2020btl}. The
exchanged particles are some light mesons such as $\pi$, $\eta$,
$\rho$ and $\omega$. In Ref. \cite{Ding:2008gr} the authors
indicated that $\sigma$ exchange makes an insignificant contribution,
thus we omit it.
The B-S equation is solved in momentum space so the kernel we
obtain by calculating the corresponding Feynman diagrams can be
used directly rather than converting it into a potential form in
coordinate space.

In this work we are only concerned about the the ground states where the orbital
angular momentum between two constituent mesons is zero (i.e. $l=0$). For
the molecular states which consist of two vector mesons their
$J^{P}$ may be $0^+$, $1^+$ and $2^+$.
For $\bar D^*K^*$ system there are four isospin states:
$\frac{1}{\sqrt{2}}(K^{*+}D^{*-}-K^{*0}\bar D^{*0})\,(I=0)$ and
$\frac{1}{\sqrt{2}}(K^{*+}D^{*-}+K^{*0}\bar D^{*0})\, (I=1, I_3=0)$,
$K^{*0}D^{*-} \,(I=1, I_3=-1)$, $K^{*+}\bar D^{*0}\, (I=1, I_3=1)$. For $
D^*K^*$ system there are also four isospin states: $K^{*+}D^{*+}\,
(I=1,I_3=1)$, $\frac{1}{\sqrt{2}}(K^{*+}D^{*0}-K^{*0}D^{*+}) \,(I=1,
I_3=0)$, $K^{*0}D^{*0}\,(I=1, I_3=-1)$ and
$\frac{1}{\sqrt{2}}(K^{*+}D^{*0}+K^{*0}D^{*+}) \,(I=0)$.

With the input parameters, these B-S equations are solved
numerically.
In some cases there no solution satisfying the
equation as long as the input parameters are  set within a
reasonable range, it implies the proposed bound state should not
appear in the nature. On the contrary,  a solution of the B-S
equation with reasonable parameters implies that the attraction
between two constituents is sufficient to build up a bound state, thus a
hadronic state (molecular) should be formed and emerge in the nature.

After this introduction we deduce the B-S equations and the
corresponding kernels for the two vector systems with different
quantum numbers. Then in section III we present our numerical
results along with explicitly displaying all input parameters.
Section IV is devoted to a brief summary.

\section{The Bethe-Salpeter formalism}

In our early papers \cite{Ke:2012gm,Ke:2020eba,Ke:2021rxd} we
deduced the B-S equations for the systems containing two vectors.
It is noted that the structures of the effective vertices are
not the same as that given in \cite{Ke:2012gm,Ke:2020eba,Ke:2021rxd},
so that the new scenario needs to be reformulated.

The quantum number $J^P$ of the bound state
composed of two vectors can be $0^+$, $1^+$ or $2^+$.

\subsection{The B-S equation of the $0^+$ system}

\begin{center}
\begin{figure}[htb]
\begin{tabular}{cc}
\scalebox{0.5}{\includegraphics{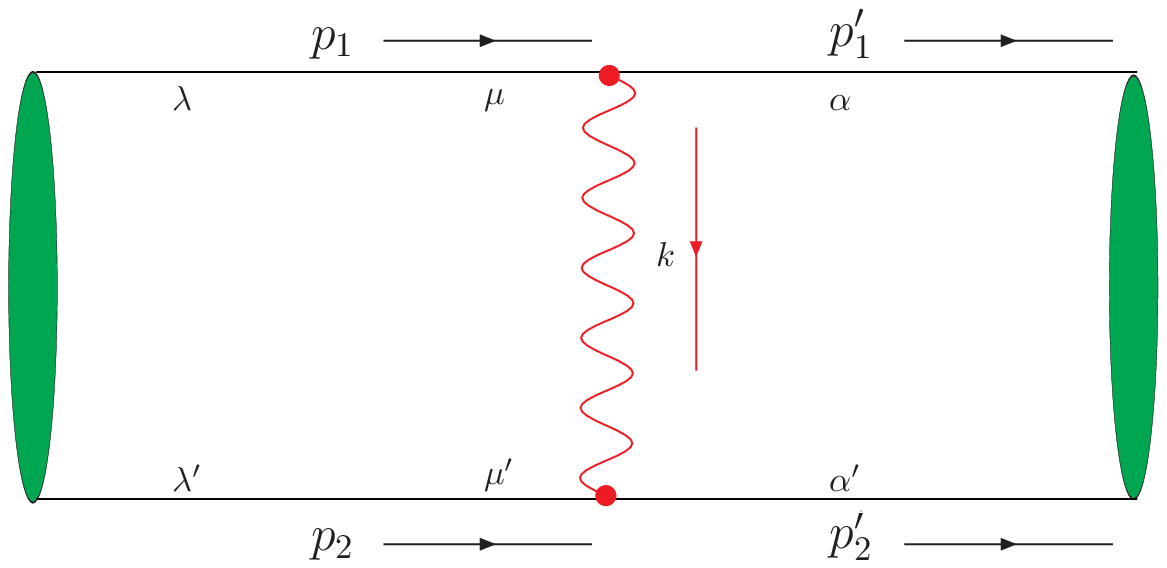}}\,\,\,\,\,\,\,\,\,\,\,\,\,\,\,\,\,\,\,\,\scalebox{0.5}{\includegraphics{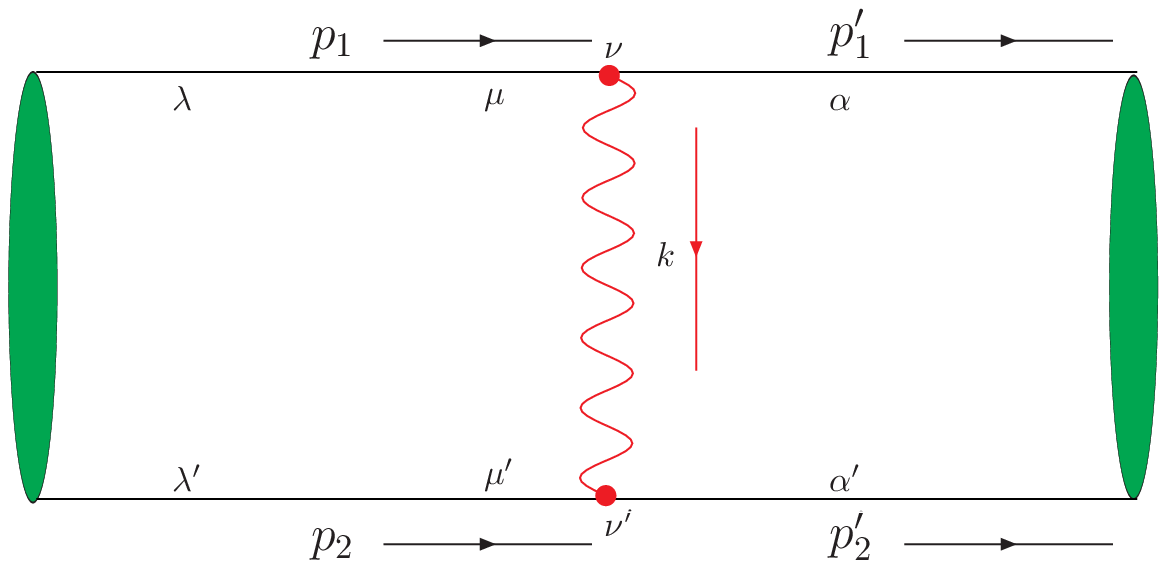}}
\\
(a)\,\,\,\,\,\,\,\,\,\,\,\,\,\,\,\,\,\,\,\,\,\,\,\,\,\,\,\,\,\,\,\,\,\,\,
\,\,\,\,\,\,\,\,\,\,\,\,\,\,\,\,\,\,\,\,\,\,\,\,\,\,\,\,\,\,\,\,\,\,\,\,
\,\,\,\,\,\,\,\,\,\,\,\,\,\,\,\,\,\,\,\,\,\,\,\,\,\,\, (b)
\end{tabular}
\caption{A bound state composed of two vectors. (a) $\pi$ ($\eta$)
is exchanged. (b) $\rho$ ($\omega$) is exchanged. }\label{DM24}
\end{figure}
\end{center}

The B-S wave function of $0^+$ state $|{S}\rangle$ composed of two
vectors is defined as following:
\begin{eqnarray}\label{definition-BS}  \langle 0 | {\rm
T}\,\phi_{1}^\mu(x_1)\phi^\nu_2(x_2) |{S} \rangle =
{\chi}_{{}_{{S}}}^{}(x_1,x_2)g^{\mu\nu}\,.
\end{eqnarray}
where $\phi^\mu_1(x_1)$ and $\phi^\nu_2(x_2)$ are the field
operators of two mesons, respectively. The relative coordinate $x$
and the center of mass coordinate $X$ are
\begin{eqnarray} X=\eta_1 x_1 + \eta_2 x_2\,,\quad x = x_1 -
x_2 \,, \end{eqnarray} where $\eta_i = m_i/(m_1+m_2)$ and $m_i\,
(i=1,2)$ is the mass of the $i$-th constituent meson.

After some manipulations we obtain the B-S equation in the
momentum space
\begin{eqnarray} \label{4-dim-BS20}
\chi_{_{{}_{S}}}({
p})=\frac{1}{4}\Delta_{1\mu\lambda}\int{d^4{p}'\over(2\pi)^4}\,K_0^{\alpha\alpha'\mu\mu'}({
p},{ p}')\chi_{_{{}_{S}}}^{}({
p}')\Delta_{2\mu'\lambda'}g_{\alpha\alpha'}g^{\lambda\lambda'},
\end{eqnarray}
where
$\Delta_{j\mu\lambda}=\frac{i}{p_j^2-m_j^2}(\frac{p_{j\mu}p_{j\lambda}}{m_j^2}-g_{\mu\lambda})$.

The relative momenta and the total momentum of the bound state in
the equation are defined as
\begin{eqnarray} p = \eta_2p_1 -
\eta_1p_2\,,\quad p' = \eta_2p'_1 - \eta_1p'_2\,,\quad P = p_1 +
p_2 = p'_1 + p'_2 \, \label{momentum-transform1}
\end{eqnarray}
where $P$ denotes the total momentum of the bound
state.

In this paper the exchanged mesons we consider  are $\pi$, $\eta$,
$\rho$ and $\omega$ \cite{Guo:2007mm,Feng:2011zzb}, and the
Feynman diagrams corresponding to these effective interactions are
depicted in Fig. \ref{DM24}.

With the Feynman diagrams and the
effective interaction we obtain
\begin{eqnarray}\label{k0}
  K_{0}^{\alpha\alpha'\mu\mu'}(p,p')&&=K_{0V}^{\alpha\alpha'\mu\mu'}(p,p',m_\rho)+
  K_{0V}^{\alpha\alpha'\mu\mu'}(p,p',m_\omega)+
  K_{0P}^{\alpha\alpha'\mu\mu'}(p,p',m_\pi)+K_{0P}^{\alpha\alpha'\mu\mu'}(p,p',m_\eta),
  \nonumber\\K_{0V}^{\alpha\alpha'\mu\mu'}(p,p',m_V)&&=iC_{V}\frac{{q_\nu
q_{\nu'}}/{{m_V}^2}-g_{\nu{\nu'}}}{q^2-m_{V}^2}[g_{_{\bar D^*\bar
D^*V}}g^{\alpha\mu} (p_1+p_1')^\nu+2g_{_{\bar D^*\bar
D^*V}}'({q}^{\alpha} g^{\mu\nu}-{q}^\mu
g^{\alpha\nu})]\nonumber\\&&g_{_{K^*K^*V}}[-q^{\alpha'}g^{\nu'\mu'}+q^{\mu'}g^{\nu'\alpha'}-{p'_2}^{\nu'}g^{\mu'\alpha'}+{p'_2}^{\mu'}g^{\nu'\alpha'}
-{p_2}^{\nu'}g^{\mu'\alpha'}+{p_2}^{\alpha'}g^{\nu'\mu'}]F(q)^2,\nonumber\\K_{0P}^{\alpha\alpha'\mu\mu'}(p,p',m_P)&&=C_{P}[g_{_{\bar
D^*\bar D^*P}}\varepsilon^{\alpha\beta\mu\nu}q_\nu
(p_1+p_1')_\beta][
-g_{_{K^*K^*P}}\varepsilon^{\alpha'\beta'\mu'\nu'}q_{\nu'}
(p_2+p_2')_{\beta'}]\frac{-i}{q^2-m_P^2}F(q)^2,
\end{eqnarray}
where $P$ and $V$ represent pseudoscalar and vector mesons. The
contributions from vector-exchanges are included in
$K_{0V}^{\alpha\alpha'\mu\mu'}(p,p',m_V)$  and those for
exchanging pseudoscalars are included in
$K_{0P}^{\alpha\alpha'\mu\mu'}(p,p',m_P)$. The isospin
coefficients  $C_{V}$ and  $C_{P}$ are collected in Tab.
\ref{iforctor1} and \ref{iforctor2} .

Since the constituent meson is not a point particle,
a form factor at each interaction vertex among hadrons must be
introduced to reflect the finite-size effects of these hadrons.
The form factor is assumed to be in the following form:
\begin{eqnarray} \label{form-factor} F({k}) = {\Lambda^2 -
M_{ V}^2 \over \Lambda^2 -{k}^2}\,,
\end{eqnarray} where $\Lambda$ is a cutoff parameter.

It is not an easy task to solve Eq.(\ref{4-dim-BS20}).
Defining $ K_{0}(p,p')=\frac{1}{4}{ K_0^{\alpha\alpha'\mu\mu'}}({
p},{ p}')(\frac{p_{2\mu'}
p_{2\lambda'}}{m_2^2}-g_{\mu'\lambda'})(\frac{{p_{1}}_\mu
{p_{1}}_{\lambda}}{m_1^2}-g_{\mu\lambda})g_{\alpha\alpha'}g^{\lambda\lambda'}$
and using the so-called instantaneous approximation: $p_0'=p_0=0$
for ${K_{0}}(p,p')$ the B-S equation can be reduced to
\begin{eqnarray} \label{3-dim-BS1}
{E^2-(E_1+E_2)^2\over (E_1+E_2)/E_1E_2}
\mathcal{\psi}_{{}_S}^{}({\bf p}) ={i\over
2}\int{d^3\mathbf{p}'\over(2\pi)^3}\, {\overline{} K_{{0}}}({\bf
p},{\bf p}')\mathcal{\psi}_{{}_S}^{}({\bf p}')\,,
\end{eqnarray}
where $E_i \equiv \sqrt{{\bf p}^2 + m_i^2}$, $E=P^0$, the
equal-time wave function is defined as $ \psi_{_S}({\bf p})= \int
dp^0 \, \chi_{_S}(p) \, $ and
\begin{eqnarray}
K_{0}(\mathbf{p},\mathbf{p}')&&=K_{0V}(\mathbf{p},\mathbf{p}',m_\rho)+{
K_{0V}}(\mathbf{p},\mathbf{p}',m_\omega)+{
K_{0P}}(\mathbf{p},\mathbf{p}',m_\pi)+{
K_{0P}}(\mathbf{p},\mathbf{p}',m_\eta).
\end{eqnarray}
The expressions of $K_{0V}(\mathbf{p},\mathbf{p}',m_V)$ and ${
K_{0P}}(\mathbf{p},\mathbf{p}',m_P)$ can be found  in Appendix B.

\subsection{The B-S equation of the $1^+$ system}

The B-S wave function of $1^+$ state $|{V}\rangle$ composed of two
vectors is defined as
 \begin{eqnarray} \label{4-dim-BS22}
\langle0|T\phi_\alpha(x_1)\phi_{\alpha'}(x_2)|{V}\rangle=\frac{\epsilon_{\alpha\alpha'\tau\tau'}}{\sqrt{6}M}\chi_{_{{V}}}(x_1,x_2)\varepsilon^\tau
P^{\tau'} ,
\end{eqnarray}
where $\varepsilon^\tau$ is the polarization vector of $1^+$
state.

The corresponding B-S equation is
\begin{eqnarray} \label{4-dim-BS42}
\chi_{_{{V}}}(p) =\frac{1}{6M^2}\epsilon^{\lambda \lambda'
\omega\sigma}\varepsilon_\sigma
P_\omega\Delta_{1\mu\lambda}\int{d^4{p'}\over(2\pi)^4}\,K_1^{\alpha\alpha'\mu\mu'}(p,p')
\epsilon_{\alpha\alpha'\omega'\sigma'}\chi_{_{{V}}}(p')\varepsilon^{\sigma'}P^{\omega'}\Delta_{2\mu'\lambda'}\,,
\end{eqnarray}
where $K_1^{\alpha\alpha'\mu\mu'}(p,p')$ is the same as $K_0^{\alpha\alpha'\mu\mu'}(p,p')$ in Eq.(\ref{k0}).

Defining $ K_{1}(p,p')=\frac{{
K_1^{\alpha\alpha'\mu\mu'}}({ p},{
p}')}{6M^2}\epsilon^{\lambda \lambda'
\omega\sigma}\varepsilon_\sigma P_\omega(\frac{p_{2\mu'}
p_{2\lambda'}}{m_2^2}-g_{\mu'\lambda'})(\frac{{p_{1}}_\mu
{p_{1}}_{\lambda}}{m_1^2}-g_{\mu\lambda})\epsilon_{\alpha\alpha'\omega'\sigma'}\varepsilon^{\sigma'}P^{\omega'}$
the B-S equation is reduced to
\begin{eqnarray} \label{3-dim-BS4}
{E^2-(E_1+E_2)^2\over (E_1+E_2)/E_1E_2}
\mathcal{\psi}_{{}_{V}}^{}({\bf p}) ={i\over
2}\int{d^3\mathbf{p}'\over(2\pi)^3}\, {\overline{} K_{1}}({\bf
p},{\bf p}')\mathcal{\psi}_{{}_{V}}^{}({\bf p}')\,,
\end{eqnarray}
with
\begin{eqnarray}
K_{1}(\mathbf{p},\mathbf{p}')&&=K_{1V}(\mathbf{p},\mathbf{p}',m_\rho)+{
K_{1V}}(\mathbf{p},\mathbf{p}',m_\omega)+{
K_{1P}}(\mathbf{p},\mathbf{p}',m_\pi)+{
K_{1P}}(\mathbf{p},\mathbf{p}',m_\eta).
\end{eqnarray}
The expressions of $K_{1V}(\mathbf{p},\mathbf{p}',m_V)$ and ${
K_{1P}}(\mathbf{p},\mathbf{p}',m_P)$ can be found  in Appendix B.

\subsection{The B-S equation of the $2^+$ system}

The B-S wave-function of $2^+$ state composed of two vectors is written as
 \begin{eqnarray} \label{4-dim-BS23}
\langle0|T\phi^\alpha(x_1)\phi^{\alpha'}(x_2)|\mathcal{T}\rangle=\frac{1}{\sqrt{5}}\chi_{_{\mathcal{T}}}(x_1,x_2)\varepsilon^{\alpha\alpha'}
,
\end{eqnarray}
where $\varepsilon^{\alpha\alpha'}$ is the polarization tensor of
the $2^+$ state.

The B-S equation can be expressed as
\begin{eqnarray} \label{4-dim-BS43}
\chi_{_{\mathcal{T}}}(p)
=\frac{1}{5}\varepsilon^{\lambda\lambda'}\Delta_{1\mu\lambda}\int{d^4{p'}\over(2\pi)^4}\,K_2^{\alpha\alpha'\mu\mu'}(p,p')
\varepsilon_{\alpha\alpha'}\chi_{_{\mathcal{T}}}(p')\Delta_{2\mu'\lambda'}\,,
\end{eqnarray}
where $K_2^{\alpha\alpha'\mu\mu'}(p,p')$ is the same as $K_0^{\alpha\alpha'\mu\mu'}(p,p')$ in Eq.(\ref{k0}).

Defining $K_{2}(p,p')= \frac{{
K_2^{\alpha\alpha'\mu\mu'}}({ p},{
p}')}{5}\varepsilon^{\lambda \lambda'
}(\frac{p_{2\mu'}
p_{2\lambda'}}{m_2^2}-g_{\mu'\lambda'})(\frac{{p_{1}}_\mu
{p_{1}}_{\lambda}}{m_1^2}-g_{\mu\lambda})\varepsilon_{\alpha\alpha'}$
the B-S equation can be reduced to
\begin{eqnarray} \label{3-dim-BS5}
{E^2-(E_1+E_2)^2\over (E_1+E_2)/E_1E_2}
\mathcal{\psi}_{{}_{T}}^{}({\bf p}) ={i\over
2}\int{d^3\mathbf{p}'\over(2\pi)^3}\, {\overline{} K_{2}}({\bf
p},{\bf p}')\mathcal{\psi}_{{}_{T}}^{}({\bf p}')\,,
\end{eqnarray}
where
\begin{eqnarray}
K_{2}(\mathbf{p},\mathbf{p}')&&=K_{2V}(\mathbf{p},\mathbf{p}',m_\rho)+{
K_{2V}}(\mathbf{p},\mathbf{p}',m_\omega)+{
K_{2P}}(\mathbf{p},\mathbf{p}',m_\pi)+{
K_{2P}}(\mathbf{p},\mathbf{p}',m_\eta).
\end{eqnarray}
The expressions of $K_{2V}(\mathbf{p},\mathbf{p}',m_V)$ and ${
K_{2P}}(\mathbf{p},\mathbf{p}',m_P)$ can be found  in Appendix B.

\section{Numerical results}
Now let us solve the B-S equations  (\ref{3-dim-BS1}),
 (\ref{3-dim-BS4}) and
(\ref{3-dim-BS5}). Since we are interested in the ground state of a bound state, the function $\psi_{J}(\mathbf{p})$ ($J$ stands for $ 0, 1$ or
$2$) only depends on
the norm of the three-momentum and we may first integrate over the
azimuthal angle of the functions in (\ref{3-dim-BS1}),
 (\ref{3-dim-BS4}) or
(\ref{3-dim-BS5})
$$\frac{i}{2}\int{d^3\mathbf{p}'\over(2\pi)^3}\, {\overline{} K_J}({\bf p},{\bf
p}'),  $$  to obtain a potential form
$U_J(|\mathbf{p}|,|\mathbf{p}'|)$ , then the B-S equation turns into a one-dimension integral
equation
\begin{eqnarray} \label{3-dim-BS6}
\psi_J({\bf |p|}) ={(E_1+E_2)/E_1E_2\over E^2-(E_1+E_2)^2 }\int{d
\mathbf{|p}'|}\, {\overline{} U_J}({\bf |p|},{\bf
|p}'|)\psi_J({\bf |p}'|) .
\end{eqnarray}
As long as the potential $U_{J}(|\mathbf{p}|,|\mathbf{p}'|)$  is attractive
and strong enough, the corresponding B-S equation has a solution(s)
and we can obtain the mass of the possible bound state. We
explained how to solve the one-dimension integral equation in Ref. \cite{Ke:2022vsi,Ke:2018jql,Ke:2021rxd,Ke:2012gm,Ke:2020eba}

\subsection{The results of $\bar D^{*}K^{*}$ system}

In our calculation the values of the parameters $g_{_{\bar D^*\bar D^*P}},
g_{_{K^*K^*P}}, g_{_{\bar D^*\bar D^*V}}$, $g'_{_{\bar D^*\bar D^*V}}$ and
$g_{_{K^*K^*V}}$ are presented in Appendix A.
 In Ref. \cite{Cheng:2004ru,Meng:2007tk} the authors
suggested a relation: $\Lambda=m+\lambda \Lambda_{QCD}$ where $m$
is the mass of the exchanged meson, $\lambda$ is a number of order
$O(1)$ and $\Lambda_{QCD}=220$ MeV. The masses of the concerned
constituent mesons, $m_{\bar D^*}$ and $m_{K^*}$ are directly
taken from the databook \cite{PDG10}.

\begin{center}
\begin{figure}[htb]
\begin{tabular}{c}
\scalebox{0.8}{\includegraphics{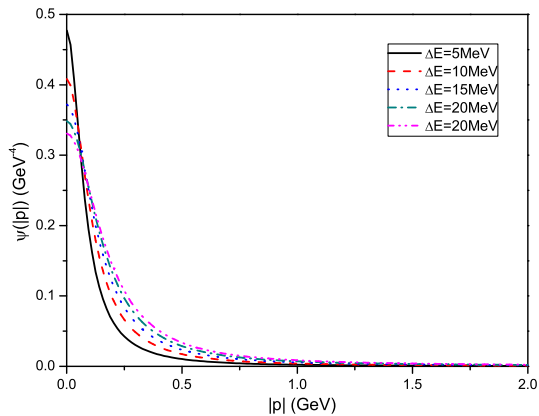}}
\end{tabular}
\caption{ the unnormalized wave functions of the bound state $\bar D^* K^*$ ($J^P=0^+$) with
 different binding energy.}\label{DM25}
\end{figure}
\end{center}

Now let us try to calculate the eigenvalues of these  systems of
$\bar D^* K^*\,(J^P=0^+, I=0)$, $\bar D^* K^*\,(J^P=1^+, I=0)$, $\bar
D^* K^*\,(J^P=2^+, I=0)$, $\bar D^* K^*\,(J^P=0^+, I=1)$, $\bar D^*
K^*\,(J^P=1^+, I=1)$ and $\bar D^* K^*\,(J^P=2^+, I=1)$ respectively.
We  find that no value of $\lambda$ can satisfy the equations of the systems
with $I=1$  i.e. these B-S equations are unsolvable. It  implies that the
effective interaction between the two constituents is
repulsive. For the system with $I=0, J=0$ or $J=1$  the B-S
equation has a solution. It implies that $\bar D^{*}K^*$ can form an
isospin scalar bound state with $J=0$ or $J=1$ by exchanging light
mesons. However the B-S equation of the system with $I=0, J=2$
still has no solution even if we vary the value of $\Lambda$ within a
larger range. It seems the attractive interaction between the two
constituents is sufficiently strong for $J=0$ or $J=1$ system but very
weak for $J=2$ one. In Ref. \cite{He:2020btl} the authors
obtained the same results with a similar approach.  In table
\ref{tab:ev1} several places are symbolized by  the mark $``-"$ which
means such bound states cannot exist due to the fact that the
corresponding B-S equation has no solution. In table \ref{tab:ev1p} the values of $\lambda$ with different binding energy $\Delta E$ are presented.
 The corresponding wave functions with different binding energy $\Delta E$ are depicted in Fig. \ref{DM25}.
 In 2020 LHCb collaboration found two
exotic states $X_0(2900)$ and $X_1(2900)$ and some theoretical
works suggested that they should be $\bar D^*K^*$ molecular
states. Our calculation supports the proposal that the two states should be
 $\bar D^*K^*$ molecular states with $I=0$. In Ref. \cite{Guo:2021mja} the authors suggested that
$X_0(2900)$ is probably compact $ud\bar s\bar c$ state with quantum number $J^P=0^+$,
but they couldn't explain the structure of $X_1(2900)$ simultaneously.

\subsection{The results for the $D^{*}K^{*}$ system}

The signs of the isospin factors $C_{\pi}$ and $C_{\rho}$ for
$D^*K^*$ system in table II are opposite to those for $\bar
D^*K^*$ shown in table I. It seems the effective interactions for $D^*K^*$
system with $I=0$ would be repulsive. However the coupling
constants $g_{_{D^*D^*V}}=-g_{_{\bar D^*\bar D^*V}}$ and
$g'_{_{D^*D^*V}}=-g'_{_{\bar D^*\bar D^*V}}$ which make the
interactions for $D^*K^*$ systems with $I=0$ still being attractive.
Indeed numerical results in table \ref{tab:ev2} support the inference i.e.
the B-S equations have solution(s) except the $J=2$ system.

For the system with $I=1$ the contributions from exchanging $\rho$
and $\omega$  nearly cancel each other and the contribution from
exchanging $\pi$ is cancelled partly by that of $\eta$ exchange so
the interaction between two constituents may be  very weak
whether it is attractive or repulsive. As we expected, at this case
these equations have no solution. Numerical results indicate these
bound states: $K^{*+}D^{*+}\, (I=1,I_3=1)$,
$\frac{1}{\sqrt{2}}(K^{*+}D^{*0}-K^{*0}D^{*+})\, (I=1, I_3=0)$ and
$K^{*0}D^{*0}\,(I=1, I_3=-1)$ should not exist but
$\frac{1}{\sqrt{2}}(K^{*+}D^{*0}+K^{*0}D^{*+})\, (I=0)$ can form
a molecular state.  Our
calculation indicates that the two
new structures $T^a_{cs0}(2900)^0$ and $T^a_{cs0}(2900)^{++}$  recently reported by the LHCb collaboration,
should not be explained as
$D^*K^*$ molecular states. Instead, in Ref. \cite{Ge:2022dsp} it is  suggested that the two peaks were resonance-like structures
induced by the threshold effects.

Moreover we find there may exist an
isoscalar $D^*K^*$ state possessing a positive charge.

\begin{table}
\caption{ The isospin factor of $\bar D^{*}K^*$ system.}
\label{iforctor1}
\begin{tabular}{c|c|c|c|c}\hline\hline
 ~~~~~~~~   &  ~~~$C_{\pi}$~~~   &
 ~~~$C_{\eta}$~~~ &  ~~~$C_{\rho}$~~~&  ~~~$C_{\omega}$~~~    \\\hline
 $I=0$    &$-\frac{3\sqrt{2}}{2}$   & $\frac{\sqrt{6}}{6}$     & $-\frac{3\sqrt{2}}{2}$  & $\frac{\sqrt{2}}{2}$  \\
 $I=1$     &$\frac{\sqrt{2}}{2}$   & $\frac{\sqrt{6}}{6}$     & $\frac{\sqrt{2}}{2}$  & $\frac{\sqrt{2}}{2}$\\
\hline\hline
\end{tabular}
\end{table}

\begin{table}
\caption{ The isospin factor of $D^{*}K^*$ system.}
\label{iforctor2}
\begin{tabular}{c|c|c|c|c}\hline\hline
 ~~~~~~~~   &  ~~~$C_{\pi}$~~~   &
 ~~~$C_{\eta}$~~~ &  ~~~$C_{\rho}$~~~&  ~~~$C_{\omega}$~~~    \\\hline
 $I=0$    &$\frac{3\sqrt{2}}{2}$   & $\frac{\sqrt{6}}{6}$     & $\frac{3\sqrt{2}}{2}$  & $\frac{\sqrt{2}}{2}$  \\
 $I=1$     &$-\frac{\sqrt{2}}{2}$   & $\frac{\sqrt{6}}{6}$     & -$\frac{\sqrt{2}}{2}$  & $\frac{\sqrt{2}}{2}$\\
\hline\hline
\end{tabular}
\end{table}

\begin{table}
\caption{ The value of $\lambda$ for $\bar D^*K^*$ system ($\Delta
E=33$ MeV).} \label{tab:ev1}
\begin{tabular}{c|c|c|c}\hline\hline
  ~~~~~~~~   &  ~~~$0^+$~~~   &
 ~~~$1^+$~~~ &  ~~~$2^+$~~~  \\\hline
 $I=0$    & 3.144  & 4.424     & -   \\
 $I=1$    & -   & -     &  -    \\
\hline\hline
\end{tabular}
\end{table}

\begin{table}
\caption{ The value of $\lambda$ for $I=0$ $\bar D^*K^*$ system
with different $\Delta E$ (in unit of MeV).} \label{tab:ev1p}
\begin{tabular}{c|c|c|c|c|c}\hline\hline
  ~~~~~~~~   &  ~~~$5$~~~   &
 ~~~$10$~~~ &  ~~~$15$~~~  &  ~~~$20$~~~ &  ~~~$25$~~~ \\\hline
 $J^P=0^+$    & 2.765  & 2.867     & 2.947 &3.008 & 3.066 \\
 $J^P=1^+$    & 3.669   & 3.868    &  4.021& 4.151&  4.265  \\
\hline\hline
\end{tabular}
\end{table}

\begin{table}
\caption{ The value of $\lambda$ for $D^*K^*$ system ($\Delta
E=33$ MeV).} \label{tab:ev2}
\begin{tabular}{c|c|c|c}\hline\hline
  ~~~~~~~~   &  ~~~$0^+$~~~   &
 ~~~$1^+$~~~ &  ~~~$2^+$~~~  \\\hline
 $I=0$    & 2.118  & 2.766     &-   \\
 $I=1$    & -   & -     &  -    \\
\hline\hline
\end{tabular}
\end{table}

\section{A brief summary}
In this work we study whether $\bar D^*K^*$  and $ D^*K^*$ can form
hadronic molecular states. We extend
our previous works where the
B-S framework was applied to the systems of  one vector and one
pseudoscalar or two vectors
\cite{Ke:2012gm,Ke:2020eba,Ke:2018jql,Ke:2021rxd,Ke:2022vsi}, to
study  possible bound states with the concerned isospin-spin structures.

In order to obtain the interaction kernels for B-S equations by
using the heavy meson chiral perturbation theory and the effective interaction to obtain the effective vertices,
we eventually are able to
calculate the corresponding Feynman diagrams where
$\pi$, $\eta$, $\rho$ or $\omega$ are exchanged.

Our final results indicate that $\bar D^*$ and $K^*$ can form two $I=0$
molecular states with $J=0$ and $J=1$. Naturally, the two states are
identified as $X_0(2900)$ and $X_1(2900)$.
As for the $D^*K^*$ system,  $I=1$
bound states  cannot exist, thus we draw a definite conclusion that $T^a_{cs0}(2900)^0$ and
$T^a_{cs0}(2900)^{++}$ cannot be explained as a $D^*K^*$
molecular state. Moreover, it is also noted that there should be an
$I=0$ $D^*K^*$ state with a positive charge.

Since the parameters are fixed from data which span a relatively
large range we cannot expect all the numerical results to be very
accurate. The goal of this work is to study whether  $\bar D^*K^*$
or  $D^*K^*$ can form a molecular state. Our results, even if not
accurate, have obvious qualitative significance. Definitely,
further theoretical and experimental works are badly needed for
gaining better understanding of these exotic hadrons.

\section*{Acknowledgments}

 This work is supported by the National Natural Science Foundation
of China (NNSFC) under the contract No. 12075167, 11975165, 11735010, 12035009 and 12075125.
\appendix

\section{The effective interactions}
The effective interactions can be found
in \cite{Colangelo:2005gb,Colangelo:2012xi,Ding:2008gr}
\begin{eqnarray}
&&\mathcal{L}_{_{D^*D^*P}}=g_{_{D^*D^*P}}(D^{*\mu}_{b}\stackrel{\leftrightarrow}{\partial}^{\beta}
D^{*\alpha\dag}_{a})(\partial^\nu
\mathcal{M})_{ba}\varepsilon_{\nu\mu\alpha\beta}+g_{_{\bar D^*\bar D^*P}}(\bar D^{*\mu}_{b}\stackrel{\leftrightarrow}{\partial}^{\beta}
\bar D^{*\alpha\dag}_{a})(\partial^\nu
\mathcal{M})_{ba}\varepsilon_{\nu\mu\alpha\beta},
\\&&\mathcal{L}_{_{D^*D^*V}}=ig_{_{D^*D^*V}}(D^{*\nu}_{b}\stackrel{\leftrightarrow}{\partial}_{\mu}
D^{*\dag}_{a\nu})(
\mathcal{V})_{ba}^\mu+ig'_{_{D^*D^*V}}(D^{*\mu}_{b}
D^{*\nu\dag}_{a}-D^{*\mu\dag}_{b} D^{*\nu}_{a})(
\partial_\mu\mathcal{V}_\nu-\partial_\nu\mathcal{V}_\mu)_{ba}\nonumber\\&&+ig_{_{\bar D^*\bar D^*V}}(\bar D^{*\nu}_{b}\stackrel{\leftrightarrow}{\partial}_{\mu}
\bar D^{*\dag}_{a\nu})(
\mathcal{V})_{ba}^\mu+ig'_{_{\bar D^*\bar D^*V}}(\bar D^{*\mu}_{b}
\bar D^{*\nu\dag}_{a}-\bar D^{*\mu\dag}_{b} \bar D^{*\nu}_{a})(
\partial_\mu\mathcal{V}_\nu-\partial_\nu\mathcal{V}_\mu)_{ba},
\end{eqnarray}
where $a$ and $b$ represent the flavors of light quarks. In
Refs. \cite{Ding:2008gr} $\mathcal{M}$ and $\mathcal{V}$ are
$3\times 3$ hermitian matrixs $
\left(\begin{array}{ccc}
        \frac{\pi^0}{\sqrt{2}}+\frac{\eta}{\sqrt{6}} &\pi^+ &K^+ \\
         \pi^- & -\frac{\pi^0}{\sqrt{2}}+\frac{\eta}{\sqrt{6}}&K^0\\
         K^-& \bar{K^0} & -\sqrt{\frac{2}{3}}\eta
      \end{array}\right)$
       and $
\left(\begin{array}{ccc}
        \frac{\rho^0}{\sqrt{2}}+\frac{\omega}{\sqrt{2}} &\rho^+ &K^{*+} \\
         \rho^- & -\frac{\rho^0}{\sqrt{2}}+\frac{\omega}{\sqrt{2}}&K^{*0}\\
         K^{*-}& \bar{K^{*0}} & \phi
      \end{array}\right)$ respectively.

In the chiral and heavy quark limit, the above coupling constants
are $g_{_{D^*D^*P}}=g_{_{\bar D^*\bar D^*P}}=\frac{g}{f_\pi},$ $g_{_{D^*D^*V}}=-g_{_{\bar D^*\bar D^*V}}=-\frac{\beta
g_V}{\sqrt{2}},\,\, g'_{_{D^*D^*V}}=-g'_{_{\bar D^*\bar D^*V}}=-\sqrt{2}\lambda g_V M_{D^*}$
with $f_\pi=132$ MeV \cite{Colangelo:2005gb},
$g=0.64$ \cite{Colangelo:2012xi}, $\kappa=g$, $\beta=0.9$,
$g_V=5.9$ \cite{Falk:1992cx} and $\lambda =0.56$ GeV$^{-1}$ \cite{Chen:2019asm}.

The effective interactions
\begin{eqnarray}
&&\mathcal{L}_{_{K^*K^*P}}=g_{_{K^*K^*P}}({\partial}^{\nu}K^{*\mu\dag}\partial^\alpha
P
K^{*\beta})\varepsilon_{\nu\mu\alpha\beta}+h.c.,
\\&&\mathcal{L}_{_{K^*K^*V}}=i\frac{g_{_{K^*K^*V}}}{2}(K^{*\mu\dag}V_{\mu\nu}
K^{*\nu}+K^{*\mu\nu\dag}V_{\mu}
K^{*}_\nu+K^{*\mu\dag}V^{\nu}
K^{*}_{\mu\nu}),
\end{eqnarray}

with $K^{*}_{\mu\nu}={\partial}_{\mu}K^{*}_{\nu}-{\partial}_{\nu} K^{*}_{\mu}$ and $V_{\mu\nu}={\partial}_{\mu}V_{\nu}-{\partial}_{\nu} V_{\mu}$.
\section{kernel}

\begin{eqnarray}
K_{0P}&&= \frac{-2 iC_{P} g_{_{\bar D^*\bar D^*P}} g_{_{K^*K^*P}} F(\bf
q)^2}{(\mathbf{p} - \mathbf{p}')^2+m_{P}^2} [-{\bf p}\cdot{\bf p}'^2+\eta_1 \eta_2 M^2 ({\bf p}-{\bf p}')^2+{\bf p}^2 {\bf p}'^2].
\end{eqnarray}

\begin{eqnarray}
&&K_{0V}=\frac{iC_{V}{g'}_{_{\bar D^*\bar D^*V}}g_{_{K^*K^*V}}
F(\bf q)^2}{[-(\mathbf{p} -  \mathbf{p}')^2-m_{V}^2]} \,\{ \frac{9\,{( {{\bf p}} - {{\bf p}'} ) }^2}{2} +
     [6\,{{{\bf p}}}^4 + {{{\bf p}}}^2\,( 2\,{{{\bf p}'}}^2 - 9\,{{\bf p}\cdot{\bf p}'} )  -
        {{M}}^2\,{{\eta_1}}^2\,( {{{\bf p}}}^2 + 2\,{{{\bf p}'}}^2 \nonumber\\&&- 3\,{{\bf p}\cdot{\bf p}'} )  + {{{\bf p}\cdot{\bf p}'}}^2 -
        4\,{{M}}^2\,{\eta_1}\,{\eta_2}\,( -{{{\bf p}}}^2 + {{\bf p}\cdot{\bf p}'} ) ]/({2\,{{m_1}}^2}) \nonumber\\&&+
     ( {{M}}^2\,{\eta_1}\,{\eta_2} + {{{\bf p}}}^2 ) \,
        [ 6\,{{{\bf p}}}^4 + {{{\bf p}}}^2\,( 2\,{{{\bf p}'}}^2 - 9\,{{\bf p}\cdot{\bf p}'} )  +
          {{M}}^2\,{\eta_1}\,{\eta_2}\,( 3\,{{{\bf p}}}^2 \nonumber\\&&+ 2\,{{{\bf p}'}}^2 - 5\,{{\bf p}\cdot{\bf p}'} )  + {{{\bf p}\cdot{\bf p}'}}^2 +
          2\,{{M}}^2\,{{\eta_2}}^2\,( -{{{\bf p}}}^2 + {{\bf p}\cdot{\bf p}'} )   ]/({2\,{{m_1}}^2\,{{m_2}}^2}) \nonumber\\&&+
   [6\,{{{\bf p}}}^4 + {{{\bf p}}}^2\,( 2\,{{{\bf p}'}}^2 - 9\,{{\bf p}\cdot{\bf p}'} )  + {{{\bf p}\cdot{\bf p}'}}^2 +
        {{M}}^2\,{{\eta_2}}^2\,( -5\,{{{\bf p}}}^2 - 2\,{{{\bf p}'}}^2 + 7\,{{\bf p}\cdot{\bf p}'} ) ]/({2\,{{m_2}}^2}) \} \nonumber\\&& +
  \frac{iC_{V}{g}_{_{\bar D^*\bar D^*V}}g_{_{K^*K^*P}}
F(\bf q)^2}{[-(\mathbf{p} -  \mathbf{p}')^2-m_{V}^2]} \,\{ {3\,{( {{{\bf p}}}^2 - {{{\bf p}'}}^2 ) }^2}/({4\,{{m_V}}^2}) +
    [\,( 12\,{{M}}^2\,{\eta_1}\,{\eta_2} +3 {( {{\bf p}} + {{\bf p}'} ) }^2 ) ]/{4}\nonumber\\&& +
     [{{{\bf p}}}^2\,{{{\bf p}'}}^2 + {{M}}^2\,{{\eta_1}}^2\,( {{{\bf p}}}^2 - {{{\bf p}'}}^2 )  +
        2\,{{M}}^2\,{\eta_1}\,{\eta_2}\,( {{{\bf p}}}^2 - {{\bf p}\cdot{\bf p}'} )  - {{{\bf p}\cdot{\bf p}'}}^2]/({4\,{{m_1}}^2})\nonumber\\&& -
     ( {{{\bf p}}}^2 - {{{\bf p}'}}^2 ) \,[ {{{\bf p}}}^2\,{{{\bf p}'}}^2 +
          {{M}}^2\,{{\eta_1}}^2\,( {{{\bf p}}}^2 - {{{\bf p}'}}^2 )  +
          2\,{{M}}^2\,{\eta_1}\,{\eta_2}\,( {{{\bf p}}}^2 - {{\bf p}\cdot{\bf p}'} )  - {{{\bf p}\cdot{\bf p}'}}^2  ]/({4\,{{m_1}}^2\,
        {{m_V}}^2})\nonumber \\&&+ [{{{\bf p}}}^2\,{{{\bf p}'}}^2 + {{M}}^2\,{{\eta_2}}^2\,( {{{\bf p}}}^2 - {{{\bf p}'}}^2 )  +
        2\,{{M}}^2\,{\eta_1}\,{\eta_2}\,( {{{\bf p}}}^2 - {{\bf p}\cdot{\bf p}'} )  - {{{\bf p}\cdot{\bf p}'}}^2]/({4\,{{m_2}}^2})\nonumber\\&& +
     ( {{{\bf p}}}^2 - {{{\bf p}'}}^2 ) \,[ {{M}}^2\,{{\eta_2}}^2\,{( {{\bf p}} - {{\bf p}'} ) }^2 -
          {{{\bf p}}}^2\,{{{\bf p}'}}^2 + {{{\bf p}\cdot{\bf p}'}}^2  ]/({4\,{{m_2}}^2\,{{m_V}}^2}) \nonumber\\&&+
   ( {{M}}^2\,{\eta_1}\,{\eta_2} + {{{\bf p}}}^2 ) \,
        [ {{{\bf p}}}^2\,{{{\bf p}'}}^2 + {{M}}^2\,{\eta_1}\,{\eta_2}\,
           ( 6\,{{{\bf p}}}^2 + {{{\bf p}'}}^2 - 3\,{{\bf p}\cdot{\bf p}'} )  +
          {{M}}^2\,{{\eta_1}}^2\,( 4\,{{{\bf p}}}^2 - 2\,{{\bf p}\cdot{\bf p}'} ) \nonumber\\&& - {{{\bf p}\cdot{\bf p}'}}^2 +
          {{M}}^2\,{{\eta_2}}^2\,( {{{\bf p}}}^2 + {{\bf p}\cdot{\bf p}'} )   ]/({4\,{{m_1}}^2\,{{m_2}}^2}) -
     ( {{M}}^2\,{\eta_1}\,{\eta_2} + {{{\bf p}}}^2 ) \,( {{{\bf p}}}^2 - {{{\bf p}'}}^2 ) \,
        [ {{{\bf p}}}^2\,{{{\bf p}'}}^2 \nonumber\\&&+ {{M}}^2\,{\eta_1}\,{\eta_2}\,( {{{\bf p}'}}^2- {{\bf p}\cdot{\bf p}'} )  -
          {{{\bf p}\cdot{\bf p}'}}^2 + {{\eta_2}}^2\,( {{M}}^2 ) (-{{{\bf p}}}^2 + {{\bf p}\cdot{\bf p}'})  ]/({4\,{{m_1}}^2\,
        {{m_2}}^2\,{{m_V}}^2}) \}.
\end{eqnarray}

\begin{eqnarray}
K_{1P}&&=\frac{-4 iC_{P} g^2_{_{D^*D^*P}} F(\bf
q)^2}{3[(\mathbf{p} - \mathbf{p}')^2+m_{P}^2]}\eta_1 \eta_2 M^2 ({\bf p}-{\bf p}')^2.
\end{eqnarray}

\begin{eqnarray}
K_{1V}&&=\frac{iC_{V}{g}_{_{\bar D^*\bar D^*V}}g_{_{K^*K^*V}}
F(\bf q)^2}{[-(\mathbf{p} -  \mathbf{p}')^2-m_{V}^2]}\,\{ {2\,{( {{\bf p}}^2 - {{{\bf p}'}}^2) }^2}/({3\,{{m_V}}^2}) +
     {2\,[ 6\,{{M}}^2\,{\eta_1}\,{\eta_2} + {( {\bf p} + {{\bf p}'}) }^2] }/{3}\nonumber \\&&+
    {( {{\bf p}}^2 - {{{\bf p}'}}^2) \,( 2\,{{\bf p}}^4 - {{\bf p}}^2\,{{{\bf p}'}}^2 - {{{\bf p}\cdot{\bf p}'}}^2) }
      /({6\,{{m_1}}^2\,{{m_V}}^2})\nonumber\\&& + {( {{\bf p}}^2 - {{{\bf p}'}}^2) \,
        [ -( {{\bf p}}^2\,{{{\bf p}'}}^2)  + {{M}}^2\,{{\eta_2}}^2\,( {{\bf p}}^2 - {{\bf p}\cdot{\bf p}'})  +
          {{{\bf p}\cdot{\bf p}'}}^2] }/({3\,{{m_2}}^2\,{{m_V}}^2}) \nonumber\\&&+
    [{{{\bf p}}^2\,{{{\bf p}'}}^2 + {{M}}^2\,{\eta_1}\,{\eta_2}\,( 6\,{{\bf p}}^2 - 4\,{{\bf p}\cdot{\bf p}'})  -
        {{{\bf p}\cdot{\bf p}'}}^2 + {{M}}^2\,{{\eta_2}}^2\,( {{\bf p}}^2 + {{\bf p}\cdot{\bf p}'}) }]/({3\,{{m_2}}^2})\nonumber\\&& +
     [{8\,{{M}}^2\,{\eta_1}\,{\eta_2}\,{{\bf p}}^2 + 2\,{{\bf p}}^4 + {{{\bf p}\cdot{\bf p}'}}^2 +
        {{\bf p}}^2\,( {{{\bf p}'}}^2 + 4\,{{\bf p}\cdot{\bf p}'}) }]/({6\,{{m_1}}^2})\}\nonumber\\&&+\frac{iC_{V}{g'}_{_{\bar D^*\bar D^*V}}g_{_{K^*K^*V}}
F(\bf q)^2}{[-(\mathbf{p} -  \mathbf{p}')^2-m_{V}^2]}\,\{ 2\,( {\bf p} - {{\bf p}'})^2 } +
     {[2\,{{\bf p}}^4 + {{\bf p}}^2\,( {{{\bf p}'}}^2 - 6\,{{\bf p}\cdot{\bf p}'})  +
        4\,{{M}}^2\,{\eta_1}\,{\eta_2}\,( {{\bf p}}^2 - {{\bf p}\cdot{\bf p}'}) \nonumber\\&& + 3\,{{{\bf p}\cdot{\bf p}'}}^2]/({3\,{{m_1}}^2}) +
     {2\,[ {{\bf p}}^2\,( 2\,{{\bf p}}^2 + {{{\bf p}'}}^2 - 3\,{{\bf p}\cdot{\bf p}'})  +
          {{M}}^2\,{{\eta_2}}^2\,( -{{\bf p}}^2 + {{\bf p}\cdot{\bf p}'}) ] }/({3\,{{m_2}}^2}\}).
\end{eqnarray}

\begin{eqnarray}
K_{2V}&&=\frac{iC_{V}{g}_{_{\bar D^*\bar D^*V}}g_{_{K^*K^*V}}
F(\bf q)^2}{[-(\mathbf{p} -  \mathbf{p}')^2-m_{V}^2]}\,\{ {2\,{( {{\bf p}}^2 - {{{\bf p}'}}^2) }^2}/({3\,{{m_V}}^2}) +
    {2\,[ 6\,{{M}}^2\,{\eta_1}\,{\eta_2} + {( {\bf p} + {{\bf p}'}) }^2] }/
      {3} \nonumber\\&&+ {( {{\bf p}}^2 - {{{\bf p}'}}^2) \,
        ( 6\,{{\bf p}}^4 - 7\,{{\bf p}}^2\,{{{\bf p}'}}^2 + {{{\bf p}\cdot{\bf p}'}}^2) }/({30\,{{m_1}}^2\,
        {{m_V}}^2})\nonumber\\&& + {( {{\bf p}}^2 - {{{\bf p}'}}^2) \,
        [ -( {{\bf p}}^2\,{{{\bf p}'}}^2)  +
          {{M}}^2\,{{\eta_2}}^2\,( {{\bf p}}^2 - {{\bf p}\cdot{\bf p}'})  + {{{\bf p}\cdot{\bf p}'}}^2] }
       /( {3\,{{m_2}}^2\,{{m_V}}^2}) \nonumber\\&&+ {2\,{{\bf p}}^2\,( {{\bf p}}^2 - {{{\bf p}'}}^2) \,
        [ -( {{\bf p}}^2\,{{{\bf p}'}}^2)  +
          {{M}}^2\,{{\eta_2}}^2\,( {{\bf p}}^2 - {{\bf p}\cdot{\bf p}'})  + {{{\bf p}\cdot{\bf p}'}}^2] }
       /( {15\,{{m_1}}^2\,{{m_2}}^2\,{{m_V}}^2})\nonumber\\&& +
    [{{{\bf p}}^2\,{{{\bf p}'}}^2 + {{M}}^2\,{\eta_1}\,{\eta_2}\,
         ( 6\,{{\bf p}}^2 - 4\,{{\bf p}\cdot{\bf p}'})  - {{{\bf p}\cdot{\bf p}'}}^2 +
        {{M}}^2\,{{\eta_2}}^2\,( {{\bf p}}^2 + {{\bf p}\cdot{\bf p}'}) }]/{3\,{{m_2}}^2} \nonumber\\&&+
     {2\,{{\bf p}}^2\,[ {{\bf p}}^2\,{{{\bf p}'}}^2 +
          {{M}}^2\,{\eta_1}\,{\eta_2}\,( 6\,{{\bf p}}^2 - 4\,{{\bf p}\cdot{\bf p}'})  -
          {{{\bf p}\cdot{\bf p}'}}^2 + {{M}}^2\,{{\eta_2}}^2\,( {{\bf p}}^2 + {{\bf p}\cdot{\bf p}'}) ] }/
       ( {15\,{{m_1}}^2\,{{m_2}}^2})\nonumber\\&& + [{40\,{{M}}^2\,{\eta_1}\,{\eta_2}\,{{\bf p}}^2 +
        6\,{{\bf p}}^4 - {{{\bf p}\cdot{\bf p}'}}^2 + {{\bf p}}^2\,( 7\,{{{\bf p}'}}^2 + 12\,{{\bf p}\cdot{\bf p}'}) }]/
       ( {30\,{{m_1}}^2})\} \nonumber\\&& + \frac{iC_{V}{g'}_{_{\bar D^*\bar D^*V}}g_{_{K^*K^*V}}
F(\bf q)^2}{[-(\mathbf{p} -  \mathbf{p}')^2-m_{V}^2]}\,
   \{ -2\,{( {\bf p} - {{\bf p}'}) }^2-\nonumber\\&&
    {2\,[ {{\bf p}}^2\,( 2\,{{\bf p}}^2 + {{{\bf p}'}}^2 - 3\,{{\bf p}\cdot{\bf p}'})  +
          {{M}}^2\,{{\eta_2}}^2\,( -{{\bf p}}^2+ {{\bf p}\cdot{\bf p}'}) ] }/({3\,{{m_2}}^2}) \nonumber\\&&+
   [{-14\,{{\bf p}}^4 + 20\,{{M}}^2\,{\eta_1}\,{\eta_2}\,
         ( {{\bf p}}^2 - {{\bf p}\cdot{\bf p}'})  - 9\,{{{\bf p}\cdot{\bf p}'}}^2 +
        {{\bf p}}^2\,( -7\,{{{\bf p}'}}^2 + 30\,{{\bf p}\cdot{\bf p}'}) }]/({15\,{{m_1}}^2})\nonumber\\&& -
     4\,[ {{\bf p}}^4\,( 2\,{{\bf p}}^2 + {{{\bf p}'}}^2 - 3\,{{\bf p}\cdot{\bf p}'})  +
          {{M}}^2\,{{\eta_2}}^2\,{{\bf p}}^2\,( -{{\bf p}}^2 + {{\bf p}\cdot{\bf p}'})+\nonumber\\&&
          {{M}}^2\,{\eta_1}\,{\eta_2}\,
           ( -3\,{{\bf p}}^4 - 3\,{{{\bf p}\cdot{\bf p}'}}^2 +
            \,  {{\bf p}}^2{{{\bf p}'}}^2 + 5 {{\bf p}}^2\,{{\bf p}\cdot{\bf p}'} )] /({15\,{{m_1}}^2\,
        {{m_2}}^2})\}.
\end{eqnarray}

\begin{eqnarray}
K_{2P}&&=\frac{4iC_{P} g^2_{_{D^*D^*P}} F(\bf
q)^2}{3[(\mathbf{p} - \mathbf{p}')^2+m_{P}^2]} \eta_1 \eta_2 M^2
({\bf p}-{\bf p}')^2.
\end{eqnarray}


\begin{thebibliography}{99}

\bibitem{Choi:2003ue}
  S.~K.~Choi {\it et al.}  [Belle Collaboration],
  Phys.\ Rev.\ Lett.\  {\bf 91}, 262001 (2003)
  [arXiv:hep-ex/0309032].

\bibitem{Abe:2007jn}
  K.~Abe {\it et al.}  [Belle Collaboration],
  Phys.\ Rev.\ Lett.\  {\bf 98}, 082001 (2007)
  [arXiv:hep-ex/0507019].


\bibitem{Choi:2005} S.~K.~Choi {\it et al.}  [Belle Collaboration],
 Phys.\ Rev.\ Lett.\  {\bf 94}, 182002 (2005).

\bibitem{Choi:2007wga}
  S.~K.~Choi {\it et al.}  [BELLE Collaboration],
  Phys.\ Rev.\ Lett.\  {\bf 100}, 142001 (2008)
  [arXiv:0708.1790 [hep-ex]].
\bibitem{LHCb:2021uow}
R.~Aaij \textit{et al.} [LHCb],
Phys. Rev. Lett. \textbf{127}, no.8, 082001 (2021)
doi:10.1103/PhysRevLett.127.082001 [arXiv:2103.01803 [hep-ex]].

\bibitem{Collaboration:2011gj}
  B.~Collaboration,
  arXiv:1105.4583 [hep-ex].


\bibitem{LHCb:2021auc}
R.~Aaij \textit{et al.} [LHCb],
[arXiv:2109.01056 [hep-ex]].


\bibitem{LHCb:2021vvq}
R.~Aaij \textit{et al.} [LHCb],
[arXiv:2109.01038 [hep-ex]].



\bibitem{GellMann:1964nj}
  M.~Gell-Mann,
  Phys.\ Lett.\  {\bf 8}, 214 (1964).




\bibitem{Chen:2016spr}
H.~X.~Chen, W.~Chen, X.~Liu, Y.~R.~Liu and S.~L.~Zhu,
Rept. Prog. Phys. \textbf{80}, no.7, 076201 (2017)
doi:10.1088/1361-6633/aa6420 [arXiv:1609.08928 [hep-ph]].




\bibitem{Guo:2019twa}
  F.~K.~Guo, X.~H.~Liu and S.~Sakai,
  doi:10.1016/j.ppnp.2020.103757
  arXiv:1912.07030 [hep-ph].

\bibitem{Molina:2010tx}
R.~Molina, T.~Branz and E.~Oset,
Phys. Rev. D \textbf{82}, 014010 (2010)
doi:10.1103/PhysRevD.82.014010
[arXiv:1005.0335 [hep-ph]].

\bibitem{Molina:2020hde}
R.~Molina and E.~Oset,
Phys. Lett. B \textbf{811}, 135870 (2020)
doi:10.1016/j.physletb.2020.135870
[arXiv:2008.11171 [hep-ph]].

\bibitem{Liu:2020nil}
M.~Z.~Liu, J.~J.~Xie and L.~S.~Geng,
Phys. Rev. D \textbf{102}, no.9, 091502 (2020)
doi:10.1103/PhysRevD.102.091502
[arXiv:2008.07389 [hep-ph]].

\bibitem{He:2020btl}
J.~He and D.~Y.~Chen,
Chin. Phys. C \textbf{45}, no.6, 063102 (2021)
doi:10.1088/1674-1137/abeda8
[arXiv:2008.07782 [hep-ph]].

\bibitem{Agaev:2020nrc}
S.~S.~Agaev, K.~Azizi and H.~Sundu,
J. Phys. G \textbf{48}, no.8, 085012 (2021)
doi:10.1088/1361-6471/ac0b31
[arXiv:2008.13027 [hep-ph]].









\bibitem{Guo:1998ef}
X.~H.~Guo, A.~W.~Thomas and A.~G.~Williams,
Phys. Rev. D \textbf{59}, 116007 (1999)
doi:10.1103/PhysRevD.59.116007 [arXiv:hep-ph/9805331 [hep-ph]].

\bibitem{Li:2019ekr}
Q.~Li, C.~H.~Chang, S.~X.~Qin and G.~L.~Wang,
Chin. Phys. C \textbf{44}, no.1, 013102 (2020)
doi:10.1088/1674-1137/44/1/013102 [arXiv:1903.02282 [hep-ph]].

\bibitem{Weng:2010rb}
  M.-H.~Weng, X.-H.~Guo and A.~W.~Thomas,
  Phys.\ Rev.\ D {\bf 83}, 056006 (2011)
  doi:10.1103/PhysRevD.83.056006
  [arXiv:1012.0082 [hep-ph]].

\bibitem{Ke:2020eba}
H.~W.~Ke, X.~H.~Liu and X.~Q.~Li,
Chin. Phys. C \textbf{44}, no.9, 093104 (2020)
doi:10.1088/1674-1137/44/9/093104 [arXiv:2004.03167 [hep-ph]].



\bibitem{Ke:2012gm}
  H.~W.~Ke, X.~Q.~Li, Y.~L.~Shi, G.~L.~Wang and X.~H.~Yuan,
  JHEP {\bf 1204}, 056 (2012)
  doi:10.1007/JHEP04(2012)056
  [arXiv:1202.2178 [hep-ph]].







\bibitem{Ke:2021rxd}
H.~W.~Ke, X.~H.~Liu and X.~Q.~Li,
Eur. Phys. J. C \textbf{82}, no.2, 144 (2022)
doi:10.1140/epjc/s10052-022-10092-8
[arXiv:2112.14142 [hep-ph]].

\bibitem{Ke:2018jql}
H.~W.~Ke and X.~Q.~Li,
Eur. Phys. J. C \textbf{78}, no.5, 364 (2018)
doi:10.1140/epjc/s10052-018-5834-9
[arXiv:1801.00675 [hep-ph]].








\bibitem{Ke:2022vsi}
H.~W.~Ke and Y.~L.~Shi,
Phys. Rev. D \textbf{105}, no.11, 114019 (2022)
doi:10.1103/PhysRevD.105.114019
[arXiv:2202.13380 [hep-ph]].

\bibitem{Ke:2021iyh}
H.~W.~Ke, X.~Han, X.~H.~Liu and Y.~L.~Shi,
Eur. Phys. J. C \textbf{81}, no.5, 427 (2021)
doi:10.1140/epjc/s10052-021-09229-y [arXiv:2103.13140 [hep-ph]].


\bibitem{Ke:2019bkf}
H.~W.~Ke, M.~Li, X.~H.~Liu and X.~Q.~Li,
Phys. Rev. D \textbf{101}, no.1, 014024 (2020)
doi:10.1103/PhysRevD.101.014024 [arXiv:1909.12509 [hep-ph]].

\bibitem{Feng:2012zzf}
  G.~Q.~Feng and X.~H.~Guo,
  Phys.\ Rev.\ D {\bf 86}, 036004 (2012).
  doi:10.1103/PhysRevD.86.036004

\bibitem{Guo:2007mm}
  X.~H.~Guo and X.~H.~Wu,
  Phys.\ Rev.\  D {\bf 76} (2007) 056004
  [arXiv:0704.3105 [hep-ph]].


\bibitem{Feng:2011zzb}
  G.~Q.~Feng, Z.~X.~Xie and X.~H.~Guo,
  Phys.\ Rev.\  D {\bf 83} (2011) 016003.




\bibitem{Oller:2000ma}
J.~A.~Oller, E.~Oset and A.~Ramos,
Prog. Part. Nucl. Phys. \textbf{45}, 157-242 (2000)
doi:10.1016/S0146-6410(00)00104-6
[arXiv:hep-ph/0002193 [hep-ph]].

\bibitem{Ding:2021igr}
Z.~M.~Ding, H.~Y.~Jiang, D.~Song and J.~He,
Eur. Phys. J. C \textbf{81}, no.8, 732 (2021)
doi:10.1140/epjc/s10052-021-09534-6
[arXiv:2107.00855 [hep-ph]].


\bibitem{Ding:2008gr}
  G.~J.~Ding,
  Phys.\ Rev.\ D {\bf 79}, 014001 (2009)
  doi:10.1103/PhysRevD.79.014001
  [arXiv:0809.4818 [hep-ph]].


\bibitem{Colangelo:2005gb}
  P.~Colangelo, F.~De Fazio and R.~Ferrandes,
  Phys.\ Lett.\ B {\bf 634}, 235 (2006)
  doi:10.1016/j.physletb.2006.01.021
  [hep-ph/0511317].

\bibitem{Colangelo:2012xi}
  P.~Colangelo, F.~De Fazio, F.~Giannuzzi and S.~Nicotri,
  Phys.\ Rev.\ D {\bf 86}, 054024 (2012)
  doi:10.1103/PhysRevD.86.054024
  [arXiv:1207.6940 [hep-ph]].

\bibitem{Casalbuoni:1996pg}
R.~Casalbuoni, A.~Deandrea, N.~Di Bartolomeo, R.~Gatto,
F.~Feruglio and G.~Nardulli,
Phys. Rept. \textbf{281}, 145-238 (1997)
doi:10.1016/S0370-1573(96)00027-0 [arXiv:hep-ph/9605342 [hep-ph]].

\bibitem{Casalbuoni:1992gi}
R.~Casalbuoni, A.~Deandrea, N.~Di Bartolomeo, R.~Gatto,
F.~Feruglio and G.~Nardulli,
Phys. Lett. B \textbf{292}, 371-376 (1992)
doi:10.1016/0370-2693(92)91189-G [arXiv:hep-ph/9209248 [hep-ph]].

\bibitem{Casalbuoni:1992dx}
R.~Casalbuoni, A.~Deandrea, N.~Di Bartolomeo, R.~Gatto,
F.~Feruglio and G.~Nardulli,
Phys. Lett. B \textbf{299}, 139-150 (1993)
doi:10.1016/0370-2693(93)90895-O [arXiv:hep-ph/9211248 [hep-ph]].

\bibitem{Cheng:2004ru}
  H.~Y.~Cheng, C.~K.~Chua and A.~Soni,
  Phys.\ Rev.\ D {\bf 71}, 014030 (2005)
  doi:10.1103/PhysRevD.71.014030
  [hep-ph/0409317].

\bibitem{Meng:2007tk}
  C.~Meng and K.~T.~Chao,
  Phys.\ Rev.\ D {\bf 77}, 074003 (2008)
  doi:10.1103/PhysRevD.77.074003
  [arXiv:0712.3595 [hep-ph]].


\bibitem{PDG10}
  K.~Nakamura {\it et al.}  [Particle Data Group],
  J.\ Phys.\ G {\bf 37}, 075021 (2010).

\bibitem{Guo:2021mja}
T.~Guo, J.~Li, J.~Zhao and L.~He,
Phys. Rev. D \textbf{105}, no.5, 054018 (2022)
doi:10.1103/PhysRevD.105.054018
[arXiv:2108.06222 [hep-ph]].

\bibitem{Ge:2022dsp}
Y.~H.~Ge, X.~H.~Liu and H.~W.~Ke,
[arXiv:2207.09900 [hep-ph]].






\bibitem{Falk:1992cx}
  A.~F.~Falk and M.~E.~Luke,
  Phys.\ Lett.\ B {\bf 292}, 119 (1992)
  doi:10.1016/0370-2693(92)90618-E
  [hep-ph/9206241].





\bibitem{Chen:2019asm}
R.~Chen, Z.~F.~Sun, X.~Liu and S.~L.~Zhu,
Phys. Rev. D \textbf{100}, no.1, 011502 (2019)
doi:10.1103/PhysRevD.100.011502
[arXiv:1903.11013 [hep-ph]].




\end{thebibliography}
\end{document}